\documentclass[twocolumn]{aastex631}

\usepackage{amsmath}
\usepackage{xspace}
\usepackage{hyperref}
\usepackage{xcolor}
\usepackage{multirow}

\newcommand{\nustar}{\textit{NuSTAR}\xspace}
\newcommand{\swift}{\textit{Swift}\xspace}
\newcommand{\nicer}{\textit{NICER}\xspace}

\newcommand{\ep}{\textit{EP}\xspace}
\newcommand{\ixpe}{\textit{IXPE}\xspace}

\newcommand{\igr}{IGR~J17091--3624\xspace}

\newcommand{\xiunit}{erg\,cm\,s$^{-1}$\xspace}

\begin{document}

\title{\igr: Newly Formed Periodic Dips and Multiwavelength Activity During the 2025 Outburst}
\correspondingauthor{Yanan Wang}
\email{wangyn@bao.ac.cn}

\author[0000-0001-9576-1870]{Zikun Lin}
\affiliation{National Astronomical Observatories, Chinese Academy of Sciences, Beijing 100101, People’s Republic of China}
\affiliation{Department of Astronomy, Xiamen University, Xiamen, Fujian 361005, People’s Republic of China}

\author[0000-0003-3207-5237]{Yanan Wang}
\affiliation{National Astronomical Observatories, Chinese Academy of Sciences, Beijing 100101, People’s Republic of China}

\author[0009-0005-5669-8465]{Shuyuan Wei}
\affiliation{School of Astronomy and Space Sciences, University of Chinese Academy of Sciences, Beijing 100049, People’s Republic of China}

\author[0000-0002-3935-2666]{Yongkang Sun}
\affiliation{National Astronomical Observatories, Chinese Academy of Sciences, Beijing 100101, People’s Republic of China}
\affiliation{School of Astronomy and Space Sciences, University of Chinese Academy of Sciences, Beijing 100049, People’s Republic of China}

\author[0000-0001-9599-7285]{Long Ji}
\affiliation{School of Physics and Astronomy, Sun Yat-sen University, Zhuhai 519082, People’s Republic of China}

\author[0000-0002-1481-4676]{Samaporn Tinyanont}
\affiliation{National Astronomical Research Institute of Thailand (Public Organisation), 260 Moo 4, T. Donkaew, A. Maerim, Chiangmai, 50180 Thailand}

\author[0000-0001-9037-6180]{Meng Sun}
\affiliation{National Astronomical Observatories, Chinese Academy of Sciences, Beijing 100101, People’s Republic of China}
\affiliation{Center for Interdisciplinary Exploration and Research in Astrophysics (CIERA), 1800 Sherman, Evanston, IL 60201, USA}

\author[0000-0003-3116-5038]{Song Wang}
\affiliation{National Astronomical Observatories, Chinese Academy of Sciences, Beijing 100101, People’s Republic of China}
\affiliation{Institute for Frontiers in Astronomy and Astrophysics, Beijing Normal University, Beijing 102206, People’s Republic of China}

\author[0000-0002-3422-0074]{Diego Altamirano}
\affiliation{School of Physics and Astronomy, University of Southampton, Southampton, Hampshire SO17 1BJ, UK}

\author[0000-0002-5341-6929]{Douglas J. K. Buisson}
\affiliation{Independent}

\author[0000-0002-0096-3523]{Wenxiong Li}
\affiliation{National Astronomical Observatories, Chinese Academy of Sciences, Beijing 100101, People’s Republic of China}

\author{Qian Chen}
\affiliation{National Astronomical Observatories, Chinese Academy of Sciences, Beijing 100101, People’s Republic of China}

\author[0000-0002-2874-2706]{Jifeng Liu}
\affiliation{ National Astronomical Observatories, Chinese Academy of Sciences, Beijing 100101, People’s Republic of China}
\affiliation{School of Astronomy and Space Sciences, University of Chinese Academy of Sciences, Beijing 100049, People’s Republic of China}
\affiliation{Institute for Frontiers in Astronomy and Astrophysics, Beijing Normal University, Beijing 102206, People’s Republic of China}
\affiliation{New Cornerstone Science Laboratory, National Astronomical Observatories, Chinese Academy of Sciences, Beijing 100012, People’s Republic of China}

\author[0000-0001-5586-1017]{Shuang-Nan Zhang}
\affiliation{Key Laboratory of Particle Astrophysics, Institute of High Energy Physics, Chinese Academy of Sciences, 19B Yuquan Road, Beijing 100049, People’s Republic of China}
\affiliation{School of Astronomy and Space Sciences, University of Chinese Academy of Sciences, Beijing 100049, People’s Republic of China}

\author[0000-0002-9702-4441]{Wei Wang}
\affiliation{National Astronomical Observatories, Chinese Academy of Sciences, Beijing 100101, People’s Republic of China}

\author[0000-0003-0292-4832]{Zhen Guo}
\affiliation{Instituto de F{\'i}sica y Astronom{\'i}a, Universidad de Valpara{\'i}so, ave. Gran Breta{\~n}a, 1111, Casilla 5030, Valpara{\'i}so, Chile}
\affiliation{Millennium Institute of Astrophysics, Nuncio Monse{\~n}or Sotero Sanz 100, Of. 104, Providencia, Santiago, Chile}

\author[0000-0003-1012-8086]{Pathompong Butpan}
\affiliation{National Astronomical Research Institute of Thailand (Public Organisation), 260 Moo 4, T. Donkaew, A. Maerim, Chiangmai, 50180 Thailand}

\author[0009-0006-9630-5352]{Rungrit Anutarawiramkul}
\affiliation{National Astronomical Research Institute of Thailand (Public Organisation), 260 Moo 4, T. Donkaew, A. Maerim, Chiangmai, 50180 Thailand}

\begin{abstract}
The black hole low-mass X-ray binary (LMXB) candidate \igr experienced a hard-state-only outburst in 2025. In this paper, we show that \ixpe detected a series of intermittent X-ray dips, spanning a total interval of $\sim$1\,day. Subsequent observations with \nicer, \ep, \nustar, and \swift\ reveal that these dips recur with a period of $2.83\pm0.07$\,days and are accompanied by an increase in spectral hardness. This is the first time such quasi-periodic dipping behavior has been observed in this target since discovery. Our spectral analysis shows that the dips can be explained by obscuration from an ionized absorber characterized by an ionization parameter of $\log \xi \sim 1$--3\,\xiunit and an equivalent hydrogen column density of $N^{\rm zxipcf}_{\rm H} \sim (1$--$30) \times 10^{22}$\,cm$^{-2}$.
The periodic reappearance of the absorber is likely caused by obscuring material located in the outer accretion disk, modulated by the binary orbital period. If confirmed, this period would suggest that the donor star in \igr has deviated from the standard main-sequence evolutionary path and is likely a (partially) stripped giant. In the optical band, no significant periodicity or correlation with the X-ray dips was detected, whereas the radio counterpart exhibited a flat to steep spectrum, in contrast to the inverted spectrum typically observed during the hard state of LMXBs.

\end{abstract}

\keywords{X-ray binary --- Accretion}

\section{Introduction} \label{sec:intro}

Low-mass X-ray binaries (LMXBs) are celestial systems consisting of a neutron star or black hole in a close orbit, accreting material via Roche lobe overflow from a donor star. Periodic X-ray dips and/or eclipses, occurring on timescales of hours to days, have been observed in approximately 30 Galactic LMXBs \citep[][]{Avakyan2023}, and are widely regarded as reliable indicators of high inclination. Eclipses occur when the X-ray–emitting region is occulted by the companion star, leading to a near-total drop in X-ray intensity. In contrast, dips are typically less regular in periodicity, exhibit shallower decreases in X-ray intensity with variable profiles, and are associated with systems viewed at moderately high inclinations ($60^{\circ}$–$75^{\circ}$; \citealt{Frank1987}).

In LXMBs, dips are typically accompanied by spectral variations, often showing increased hardness ratios \citep[see e.g.,][]{White1982, Boirin2005, Trigo2006, Shidatsu2013, Buisson2021}. Accordingly, dipping phenomena are thought to result from the temporary obscuration of the central X-ray source by clumps of low-ionization material along the line of sight \citep[see e.g.,][]{Boirin2005, Trigo2006, Shidatsu2013}.
Furthermore, the relatively long recurrence timescales of these dips are generally consistent with the binary orbital period, suggesting that the obscuring material is located in the outer regions of the accretion disk and is likely associated with the mass-transfer stream from the donor star \citep[e.g.,][]{Frank1987, Armitage1996, Armitage1998}. Among dipping LMXBs, some sources (such as EXO 0748--676 \citealt{Crampton1986} and MAXI J1659--152 \citealt{CorralSantana2018}) also exhibit optical modulation at the same period as the X-ray dips, further supporting their orbital origin.
In addition, dips can show evolving profiles and may even appear early in the outburst before disappearing later (e.g., \citealt{Kuulkers2013,Kajava2019}). This behavior can be qualitatively explained by disk precession \citep{Trigo2009} or changes in the structure of the accretion disk \citep{Kuulkers2013}. Therefore, X-ray dips serve as powerful diagnostics, offering valuable insights into the orbital parameters and dynamics of X-ray binaries, as well as the structure and behavior of their accretion disks.

\igr is a Galactic low-mass black hole X-ray binary (BHXRB) candidate, first discovered by {\it INTEGRAL/IBIS} in April 2003 \citep{Kuulkers2003}. Since then, it has undergone four additional outbursts in 2007, 2011, 2016, and 2022 \citep{Capitanio2009, Krimm2011, Altamirano2011, Miller2016, Miller2022, Wang2024}. Compared to other systems, \igr displays a much richer variety of X-ray variability on timescales ranging from sub-seconds to minutes \citep{Altamirano2011, Court2017, Wang2024}, including heartbeat variability and quasi-periodic oscillations. However, system parameters, such as the black hole mass, orbital period, and the spectral type of the donor star, remain uncertain, hindering a deeper understanding of the origin of these variations.

In February 2025, \igr underwent a new outburst \citep{ATel17034,ATel17038}, which coincided with an optical re-brightening \citep{ATel17065,ATel17166}. 
Follow-up observations were triggered with several X-ray telescopes. Among them, \textit{Imaging X-ray Polarimetry Explorer} (\ixpe; \citealt{Weisskopf2022}) conducted a long observation of \igr from March 7 to 10, 2025, during which dips were detected only on approximately one day of the exposure \citep{Ewing2025}. The overlapped \textit{Nuclear Spectroscopic Telescope Array} \citep[\nustar;][]{Harrison2013} observation confirms the presence of dips \citep{Ewing2025}. Subsequent spectral analyses using \nustar observations have attributed these dips to absorption \citep{Debnath2025,Adegoke2025}. \cite{Pahari2013} had previously reported dips in \igr during its 2011 outburst, though these were distinct from absorption dips. The appearance of absorption dips in \igr in 2025 presents an opportunity to probe its binary system parameters. In this work, we present a follow-up study of the X-ray dipping phenomenon in \igr, examine its multi-wavelength emission, and discuss the properties and possible origins of these features.

\section{Observations and Data reduction} \label{sec:Methods}

\subsection{X-ray observations}
This paper used X-ray data from \textit{Neutron Star Interior Composition ExploreR} \citep[\nicer;][]{Gendreau2016}, \textit{Einstein Probe} Follow-up X-ray Telescope \citep[\ep/FXT;][]{Yuan2022}, the X-Ray Telescope on board the \textit{Neil Gehrels Swift Observatory} \citep[\swift/XRT;][]{Burrows2005}, \nustar, and \ixpe. Unless explicitly mentioned, the source region used to extract the X-ray data is centered at R.A. = 257.2817$^\circ$ and Decl. = -36.4070$^\circ$ (J2000). \ep/FXT and \swift/XRT are hereafter referred to as \ep and \swift, respectively, throughout this paper.

All X-ray spectra were rebinned using the \texttt{ftgrouppha} task \citep{Kaastra2016} from the {\sc ftools} package \citep{Blackburn1995}, with the \texttt{optmin} option to ensure a minimum of 3 counts per bin. Spectral fitting was carried out in \textsc{xspec} v12.13.1 \citep{arnaud1996} using C-statistics \citep{Cash1979,Kaastra2017}, the abundance tables and photoelectric cross-sections were taken from \cite{Wilms2000} and \cite{Verner1996}, respectively. Unless explicitly mentioned, the uncertainties for each fitting parameter in this work were calculated at 1-$\sigma$ confidence level.

\subsubsection{\nicer}

\nicer monitored \igr from March 15 to May 31, 2025 (PIs: F. Vincentelli, Z. Lin, G. Mastroserio, and F. Sanzenbacher). Raw data were download from High Energy Astrophysics Science Archive Research Center (HEASARC). Data processing and screening were performed with the \texttt{nicerl2} pipeline within \textsc{HEASoft} v6.32.1 \citep{Heasarc2014ascl.soft08004N}, applying the calibration files in version xti20221001. The lightcurves and spectra were generated with \texttt{nicerl3-lc} and \texttt{nicerl3-spect}, respectively. To minimize the impact of background flares and ensure an adequate signal-to-noise ratio (SNR), we regenerated the good time intervals (GTIs) by selecting periods with a 12--15\,keV count rate below 0.5\,counts/s and a minimum duration of 100\,seconds.
The spectrum background was estimated by SCORPEON\footnote{\href{https://heasarc.gsfc.nasa.gov/docs/nicer/analysis_threads/scorpeon-overview/}{https://heasarc.gsfc.nasa.gov/docs/nicer/analysis\_threads/scorpeon-overview/}} background model. We let the \texttt{swcx} parameters free to vary to account for the excesses below 0.6\,keV caused by the solar wind charge exchange. Spectral fitting was performed in the 0.5–15\,keV energy range to characterize the background spectrum.

\subsubsection{\textit{Einstein Probe}}

\ep monitored \igr from March 15 to May 31, 2025 (PI: Y. Wang). The raw data were processed using the FXT Data Analysis Software (FXTDAS) v1.10 with CALDB v1.10. Source light curve and spectra files for the two co-aligned instruments, FXTA and FXTB, were generated following the standard procedure provided by the FXT Data Center\footnote{\href{http://epfxt.ihep.ac.cn/analysis}{http://epfxt.ihep.ac.cn/analysis}}. The source was extracted using a circular region with a radius of 60$^{\prime\prime}$. The background region was selected using an annulus centered on the same coordinates as the source region, with inner and outer radii of 120\arcsec and 240\arcsec, respectively.

\subsubsection{\swift}

\swift monitored \igr from February 13 to June 28, 2025 (PI: S. Motta, M. Parra, and Y. Wang). The \swift light curves and spectrum were generated by the online tools at UK \swift Science Data Centre\footnote{\url{https://www.swift.ac.uk/user\_objects/}} \citep{Evans2007,Evans2009}. The observation was split into snapshots (spacecraft orbits). Observations with fewer than 100 total counts were excluded to ensure a sufficient SNR for reliable spectral fitting.

\subsubsection{\ixpe}

\ixpe observed \igr from March 7 to March 10, 2025 (PI: M. Parra). We downloaded the Level-2 event files from the HEASARC and applied barycenter corrections using the \texttt{barycorr} task from the {\sc ftools} package \citep{Blackburn1995}. Source events were extracted using the \texttt{fselect} command from a circular region with a radius of 60\arcsec, while background events were selected from an annular region with an inner radius of 120\arcsec and an outer radius of 240\arcsec. We combined the light curves from the three detector units using {\it Stingray} \citep{Huppenkothen2019}.

\subsubsection{\nustar}

\nustar took five observations of IGR~J17091 (PI: J. Garcia) between Feb 12 and March 9, 2025. The raw data was download from HEASARC, and we conducted data processing using the \nustar Data Analysis Software (NUSTARDAS) version 1.9.7 with the Calibration Database version v20230420. The data was calibrated using the command {\it nupipeline}.
We generated the source spectra and lightcurves for the two co-aligned instruments, FPMA and FPMB, in the energy range 3--79\,keV, respectively, along with their corresponding response and ancillary response files, using the {\it nuproducts} task. The extraction of the source region is circled with a radius of 100\arcsec, where the center is set at the emission peak. The background region was selected by placing a circular area with the same radius as the source region at the farthest corner of the image from the source center.

\begin{deluxetable*}{lcccccc}[!ht]
\tablecaption{Radio observations of \igr by ATCA}
\label{tab:radio_data}
\tablewidth{0pt}
\tablehead{
\colhead{MJD} & \colhead{Frequency} & \colhead{Flux$^a$} & \colhead{Beam size (major)} & \colhead{Beam size (minor)} & \colhead{rms} & \colhead{$\alpha$} \\
& (GHz) & (mJy) & (arcsec) & (arcsec) & (mJy/beam) &
}
\startdata
\multirow{4}{*}{60729.8156} & 5.0 & $1.09\pm0.03$ & 13.15 & 1.52 & 0.022 & \multirow{4}{*}{-0.16$\pm$0.02} \\
& 6.0 & $1.08\pm0.03$ & 11.37 & 1.26 & 0.019 & \\
& 8.5 & $1.01\pm0.03$ & 8.10 & 0.93 & 0.018 & \\
& 9.5 & $1.00\pm0.03$ & 6.99 & 0.80 & 0.017 & \\
\hline
\multirow{4}{*}{60734.6821} & 5.0 & $1.01\pm0.04$ & 11.30 & 1.86 & 0.023 & \multirow{4}{*}{-0.51$\pm$0.07} \\
& 6.0 & $0.90\pm0.04$ & 9.42 & 1.57 & 0.021 & \\
& 8.5 & $0.80\pm0.03$ & 6.68 & 1.12 & 0.016 & \\
& 9.5 & $0.71\pm0.03$ & 5.63 & 0.99 & 0.015 & \\
\hline
\multirow{4}{*}{60735.6355} & 5.0 & $0.88\pm0.08$ & 9.64 & 1.76 & 0.044 & \multirow{4}{*}{-0.67$\pm$0.04} \\
& 6.0 & $0.80\pm0.08$ & 8.16 & 1.48 & 0.040 & \\
& 8.5 & $0.62\pm0.06$ & 5.93 & 1.07 & 0.027 & \\
& 9.5 & $0.57\pm0.07$ & 4.88 & 0.93 & 0.025 & \\
\enddata
\tablecomments{
$^a$: Integrated flux density.\\
}
\end{deluxetable*}

\subsection{Optical observations}

Follow-up optical observations of \igr was carried out by Thai Robotic Telescope (TRT; Program ID: TRTC12A\_003 and TRTC12B\_003; PI: K. S. Tinyanont) and Southern Astrophysical Research telescope (SOAR; PI: I. El Mellah) from February 20 to June 30, 2025 using the g$^\prime$, r$^\prime$, i$^\prime$, and z$^\prime$ filters. The calibrated stars were selected within 60\arcsec~of \igr from SkyMapper Southern Survey DR4 \citep{Onken2024}. Photometry was performed using the \texttt{photutils} package. The source aperture radius was chosen to enclose 90\% of the photons, determined by averaging the widths from Gaussian profile fits in the vertical and horizontal directions.

There is an optical source located around 1\arcsec~from \igr \citep{ATel17065}. Due to the limited spatial resolution of TRT and SOAR, the two sources could not be fully resolved in our observations. As a result, our photometry includes contributions from both \igr and this nearby source. The observed magnitude during the 2025 outburst is at least one magnitude brighter than that measured when \igr was in quiescence in 2024 by the Las Cumbres Observatory (LCO) network of telescopes \citep{ATel17065}. Additionally, SOAR can partially resolve the nearby source from \igr (see Fig.~\ref{fig:SOAR_diff_image.png}), and the difference image indicates that the re-brightening is primarily dominated by \igr (see Sec.~\ref{sec:Appendix_Opt} for more details). We hence conclude that the optical variations observed in the bottom panel of Fig.~\ref{fig:lc.png} are dominated by \igr.

\subsection{Radio observations}

The Australia Telescope Compact Array (ATCA) observed \igr during three epochs: February 23, February 28, and March 1, 2025 (Project: C3615; PI: Y. Wang). All observations were obtained in the 6D array configuration, which offers the longest baseline and hence the highest spatial resolution. Two spectral windows were centred at 5.5\,GHz and 9.0\,GHz, each with a bandwidth of 2\,GHz. Absolute flux-density and band-pass calibration were performed using 1934--638 for the first epoch and 0823--500 for the subsequent two epochs, when 1934--638 was below the horizon. 1714--336 served as the phase calibrator throughout. No further ATCA observations were conducted after March 2025, as the array entered the commissioning phase of the Broadband Integrated GPU Correlator for ATCA.

We reduced the data using the Common Astronomy Software Application (CASA v6.6.5, \citealt{CASATeam2022}). After standard calibration procedures, we imaged the calibrated visibilities using the CASA task {\tt tclean} with Briggs weighting (robust = 0.5), and generated four 1 GHz-wide sub-band images centred at 5.0, 6.0, 8.5, and 9.5\,GHz to characterize the spectral evolution of the source. Flux densities were extracted with the task {\tt imfit}, fitting a single Gaussian component at the position of \igr. The quoted uncertainty is the quadrature sum of the statistical error returned by {\tt imfit} and a 5\% systematic term accounts for the absolute flux-density scale. The resulting measurements for all epochs are listed in Tab.~\ref{tab:radio_data}. The time of the ATCA observations are indicated by blue dashed lines in Fig.~\ref{fig:lc.png}.

\section{Results} \label{sec:Result}

\subsection{X-ray Timing Analysis} \label{sec:Timing_Analyize}

\begin{figure*}[!t]
 \centering
   \includegraphics[width=\linewidth]{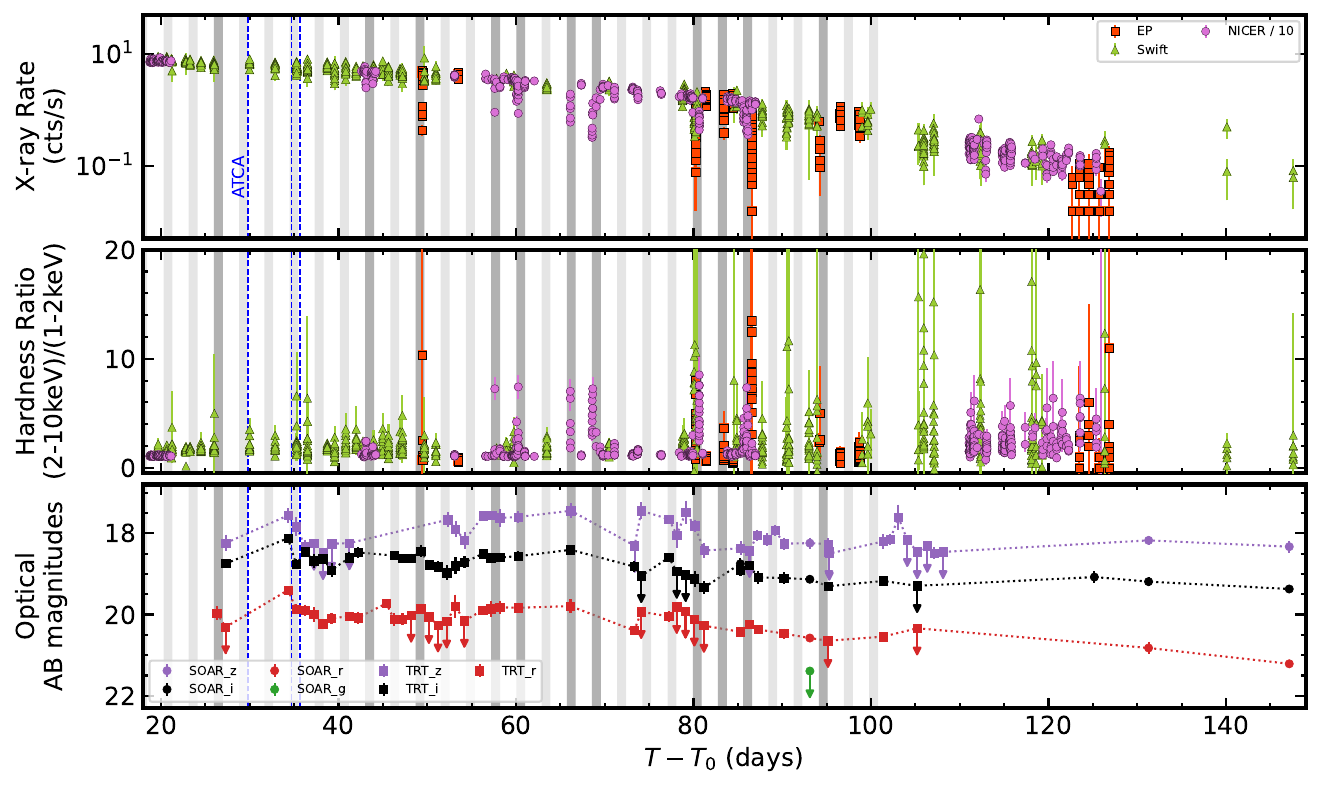}
   \caption{
   \textit{Top panel}: 
   The X-ray light curves of \igr, observed by \nicer (magenta), \ep (red), and \swift (green). The \nicer count rate was divided by 10 to align with those from the other telescopes. The light-gray shaded regions represent the predicted dipping episodes, while the darker gray regions represent the detected dipping episodes. The light curves were binned with a time resolution of 64\,s. The blue vertical dashed lines indicate the epochs of the ATCA radio observations. 
   \textit{Middle panel}: 
   The HR derived from the above observations.
   \textit{Bottom panel}: 
   The optical light curves of \igr observed by SOAR and TRT in the g$^\prime$ (green), r$^\prime$ (red), i$^\prime$ (black), and z$^\prime$ (purple) filters. The downward arrows denote $3\sigma$ upper limits.
   }
   \label{fig:lc.png}
\end{figure*}

Fig.~\ref{fig:lc.png} shows the long-term light curves and hardness ratio (HR) from \nicer, \ep, and \swift observations. The HR is defined as the hard X-ray (2--10\,keV) count rate divided by the soft X-ray (1--2\,keV) count rate. In this study, we adopted MJD 60700 as the starting point ($T_{\rm 0}$) of the dataset. Approximately 27\,days after $T_{\rm 0}$, a series of repeated dips were observed by \ixpe, \nicer, \nustar, \ep, and \swift. During these dips, an increase in HR was observed, indicating that the flux reduction was more pronounced in the soft X-rays than in the hard X-rays.

Among these observations, \nicer provided the most extensive coverage of the outburst, detecting seven prominent dip series (defined as \nicer $\rm HR > 2$) before day 100. After day 100, the X-ray emission exhibited a more rapid decline. Concurrently, the variability increased and became more irregular, accompanied by increased fluctuations in the HR. These fluctuations make it difficult to reliably identify dips in the light curve. Therefore, data collected after day 100 are excluded from the period determination.

\begin{figure}[!t]
 \centering
   \includegraphics[width=1\linewidth]{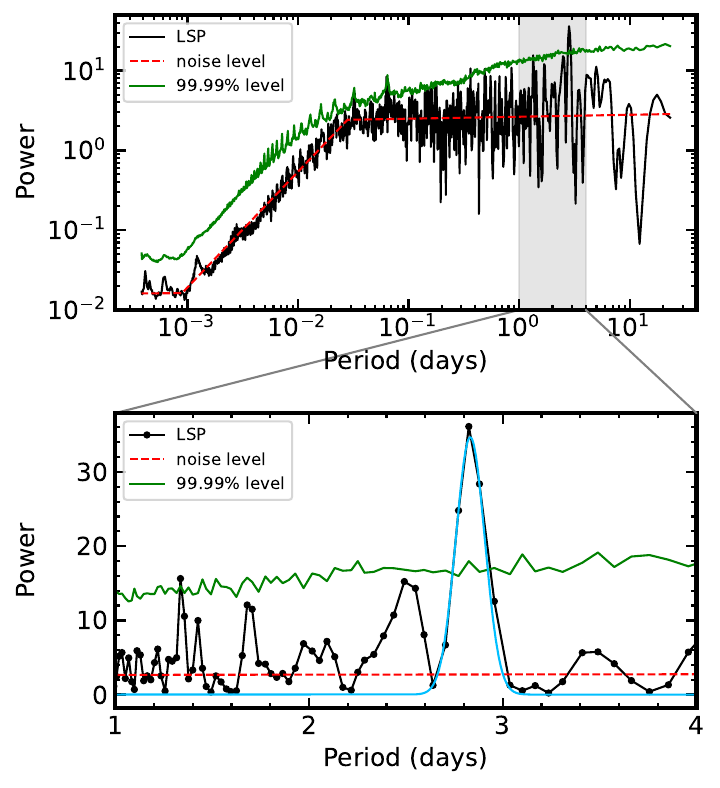}
   \caption{ 
   \textit{Top panel}: LSP of the detrended \nicer light curve. The red dashed line indicates the noise level, while the green line marks the 99.99\% (4$\sigma$) confidence level.
   \textit{Bottom panel}: Zoomed-in view of the LSP around the most prominent peak. The blue curve shows the best-fitting Gaussian profile centered at $2.83\pm0.07$\,days.
   }
   \label{fig:LSP.png}
\end{figure}

To measure the periodicity of these dips, we first detrended the \nicer soft X-ray light curve within the first 100\,days by fitting a broken exponential decay model \citep[][]{King1998,Shahbaz1998}:
\begin{align}
f(x) =
\begin{cases}
N \, \exp\!\left(-\dfrac{x - x_{\mathrm{b}}}{\tau_1}\right), & \text{if } x \le x_{\mathrm{b}} \\[8pt]
N \, \exp\!\left(-\dfrac{x - x_{\mathrm{b}}}{\tau_2}\right), & \text{if } x \ge x_{\mathrm{b}}
\end{cases}
\label{eq:broken_exp_decay}
\end{align}
where $x_{\mathrm{b}}$ denotes the break time, $\tau_1$ and $\tau_2$ are the e-folding timescales before and after the break, respectively, and $N$ is the normalization parameter. We note that this function is employed to detrend the long-term decay of the X-ray light curve; alternative forms, such as a broken linear function or a combination of exponential and linear components, can also describe the decay and do not substantially affect our subsequent analysis. The fit was performed using data with HR$<1.2$, in order to minimize the influence of the dips. The light curve was binned at 16\,s. We then applied the Lomb–Scargle periodogram (LSP) method, as implemented in the \texttt{astropy} library \citep{Astropy2022}, to the detrended light curve to search for periodic signals.
The LSP was logarithmically rebinned with a factor of 0.01. As a first approximation, we estimated the peak period and its uncertainty by fitting a Gaussian to the peak profile, yielding $2.83 \pm 0.07$\,days (Fig.~\ref{fig:LSP.png}).

To evaluate the significance of this detected periodicity, we first modeled the noise level in the LSP using a double broken power-law to account for the white and red noise components:
\begin{align}
f(x) = 
\begin{cases} 
N\left( \frac{x}{x_\mathrm{b1}} \right)^{p_1}, & \text{if } x \le x_{\rm b1} \\
N\left( \frac{x}{x_\mathrm{b1}} \right)^{p_2}, & \text{if } x_{\rm b1} \le x \le x_{\rm b2} \\
N\left( \frac{x_\mathrm{b2}}{x_\mathrm{b1}} \right)^{p_2} \left( \frac{x}{x_\mathrm{b2}} \right)^{p_3}, & \text{if } x \ge x_{\rm b2}
\end{cases}
\end{align} 
where $x_{\rm b1}$ and $x_{\rm b2}$ represent the two break frequencies, $p_1$, $p_2$, and $p_3$ denote the power-law index of the three spectral segments, and $N$ is the normalization parameter. Subsequently, we generated 50,000 simulated light curves using the DELightcurveSimulation\footnote{\url{https://github.com/samconnolly/DELightcurveSimulation}} code \citep{Connolly2016}, with the Emmanoulopoulos method \citep{Emmanoulopoulos2013}. Then, we generated the LSP of the simulated light curves and estimated their power distribution to determine the significance levels. The detected periodic signal exceeded a $99.99\%$ (4$\sigma$) confidence level based on these simulations, robustly indicating that the period is real.

\begin{figure}[!t]
 \centering
   \includegraphics[width=\linewidth]{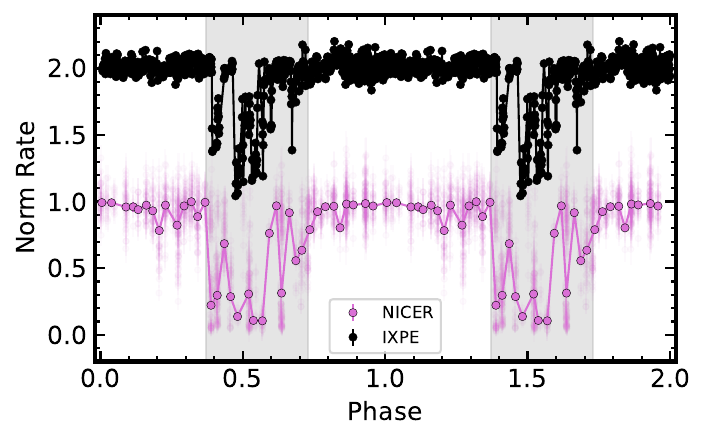}
   \caption{Folded soft X-ray light curve from \nicer (detrended), shown alongside the \ixpe light curve, using a period of 2.84\,days. The \nicer phase curve are rebinned into 0.025 phase units (corresponding to 0.07 days) using a weighted mean for clarity, with the original phase curve shown transparently. The gray shaded regions represent the dipping periods with a duration of 0.36 phase units (1\,day).
   }
   \label{fig:Phase_Xray.png}
\end{figure}

We folded the \nicer light curves over the period uncertainty range of $2.83\pm0.07$\,days, in steps of 0.01\,days, and found that a period of 2.84\,days provides the best phase alignment with the \ixpe observations. The series of dips spans approximately 1\,day, corresponding to a phase interval of 0.36 (see Fig.~\ref{fig:Phase_Xray.png}). Under the assumption that the dips exhibit no phase shift over time, we adopted 2.84\,days as the period for further analysis and used it to search for other potential dipping episodes.

Based on the above period, we identified eight dips in the \ep, \nustar, and \swift data within the first 100\,days of the outburst, four of which occurred outside \nicer coverage. In total, 11 of the 29 predicted dipping episodes were detected; apart from the very first \nicer observations (days 20.3--21.3), the remaining missed episodes were due to lack of coverage. The identification of the non-detection in this \nicer observation is based on the criterion that HR$<2$, and that spectral modeling does not statistically require an additional absorption component (see Section~\ref{sec:Spectral_ana} for details), with $\Delta$C-stat$<1$ for two additional degrees of freedom (DOF). This absence suggests that the dipping behavior may have developed gradually over the course of the outburst.

\subsection{X-ray Spectral Analysis} \label{sec:Spectral_ana}

\begin{figure*}[!ht]
 \centering
   \includegraphics[width=\linewidth]{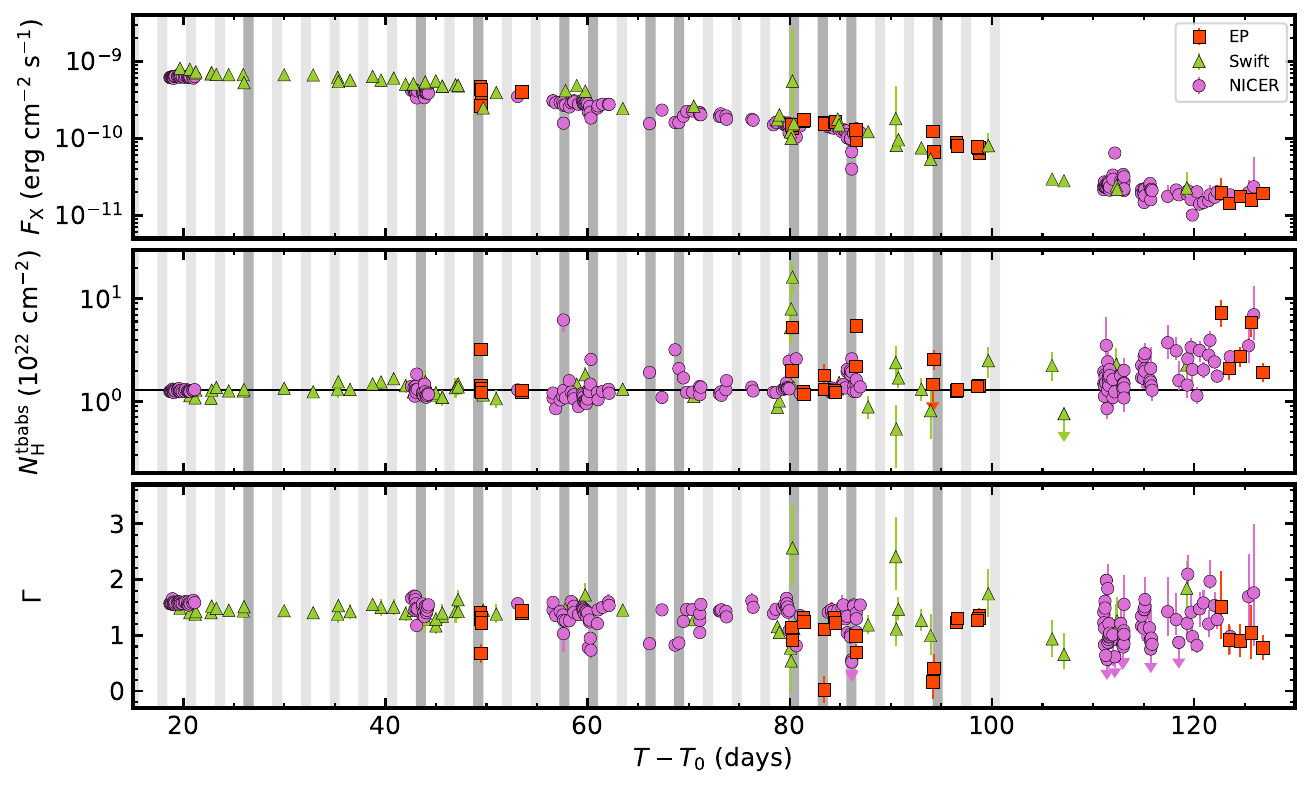}
   \caption{The evolution of spectral parameters derived from individual fitting, with the unabsorbed flux measured in the 0.5--10\,keV band. Due to a different energy coverage, we do not include the \ixpe and \nustar data in this plot. The color are defined the same as in Fig.~\ref{fig:lc.png}. The horizontal black line in middle panel indicates the median value of the $N^{\rm tbabs}_{\rm H}$ and is included to guide the eye for parameter evolution.
   }
   \label{fig:spe_all.png}
\end{figure*}

To investigate the origin of the dips, we conducted spectral analyses throughout the outburst. Initially, we modeled individual spectra on a GTI basis using an absorbed power-law model (\texttt{tbabs} $\times$ \texttt{powerlaw}). For the \ep and \nustar observations, we jointly fitted their spectrum from the two co-aligned units, using a \texttt{constant} parameter to account for calibration differences. This model provided a good fit to the continuum during the non-dipping episodes, with the photon index ($\Gamma$) remaining below 2 within uncertainties. This suggests that \igr has been in the hard state throughout the outburst. We also tested adding a disk blackbody component (\texttt{diskbb}; \citealt{Mitsuda1984}), but it did not result in significant improvement to the fit. Therefore, we applied the absorbed \texttt{powerlaw} model to describe the spectra from all detectors. The best-fitting parameters for each individual spectrum are shown in Fig.~\ref{fig:spe_all.png}.

As shown in Fig.~\ref{fig:spe_all.png}, the dipping episodes are accompanied by an increase in the column density ($N^{\rm tbabs}_{\rm H}$), indicating the presence of additional absorbing materials. Furthermore, we noticed a coupling between $N^{\rm tbabs}_{\rm H}$ and $\Gamma$, with the column density increasing as the photon index decreases. In several cases, a noticeable excess appears in the residuals around the 1--2\,keV energy band, suggesting that neutral absorption alone is insufficient to fully explain the observed dips.
After day 100, $N^{\rm tbabs}_{\rm H}$ gradually increased, suggesting an evolution of the absorbing material along the line of sight.

\begin{figure*}[!ht]
 \centering
   \includegraphics[width=\linewidth]{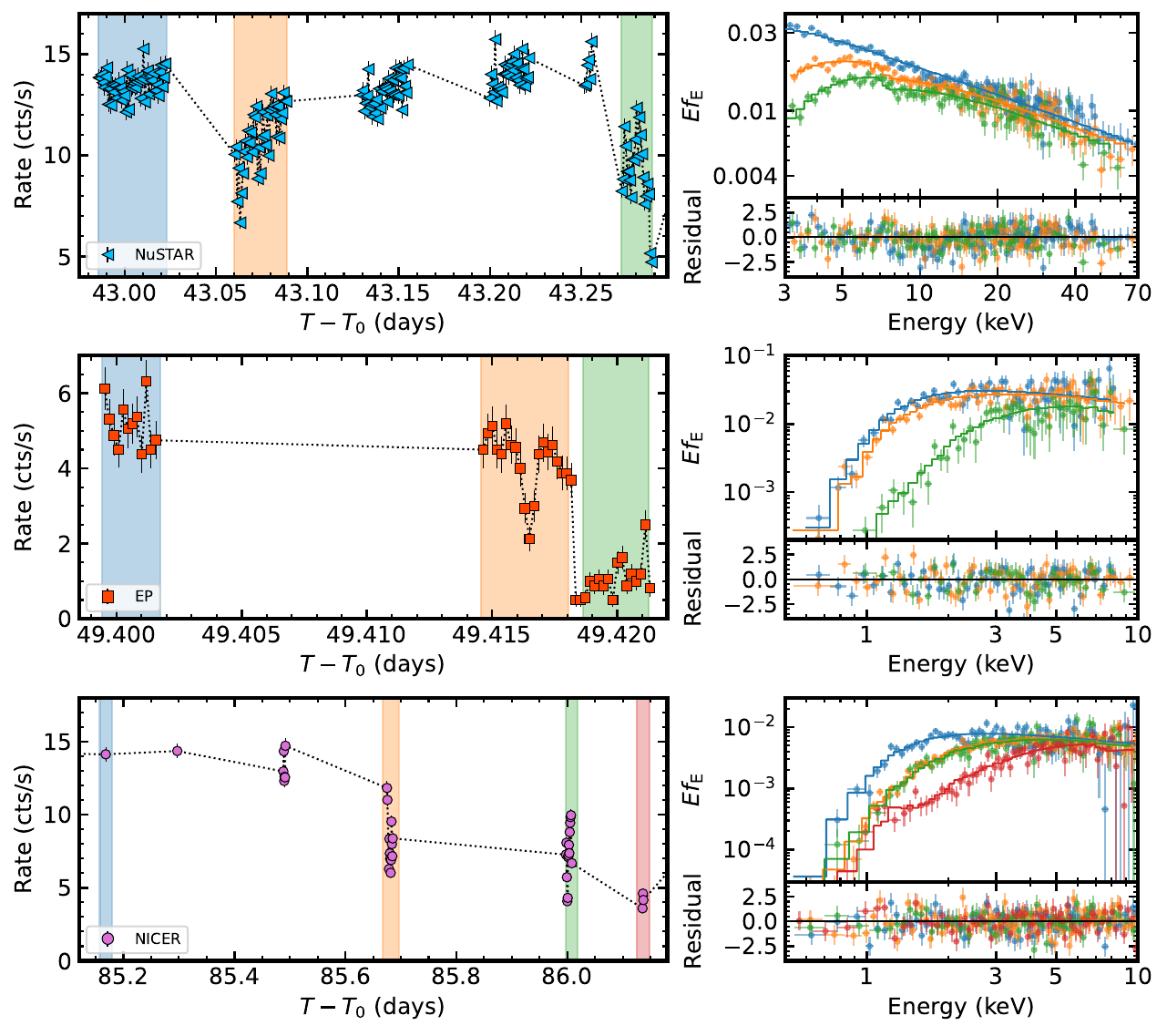}
   \caption{Three representative examples of light curve segments (left panels) used for joint spectral fitting, along with their corresponding best-fit spectra (right panels). The color of each spectrum corresponds to one GTI, as including the shaded regions of the same color in the light curves. The blue region indicates the persistent interval, while the other shaded regions indicate the dipping intervals. We note here that we only show several example spectra for clarity; the data outside the shaded regions are also included in the joint fitting.
   }
   \label{fig:SPE_plot.png}
\end{figure*}

To account for the additional absorption component, we first included an ionized absorber component, \texttt{zxipcf}, resulting in a model: \texttt{constant}$\times$\texttt{tbabs}$\times$\texttt{zxipcf}$\times$\texttt{powerlaw}. The \texttt{constant} component was used to account for calibration between different instruments. 
Moreover, to separate the ionized absorption from the continuum, we performed joint spectral fits of dipping and adjacent non-dipping episodes within the same cycle ($\pm1.4$\,days). The time coverage of each joint fit episodes is shown by the orange shaded region in the right panel of Fig.~\ref{fig:SPE_phase_evo_result.png}. 
Assuming the continuum and interstellar absorption do not vary within a cycle, we tied the \texttt{tbabs} and \texttt{powerlaw} parameters across dipping and non-dipping intervals. 
In addition, we observed a flux decrease of $3$--$15\%$ above 20\,keV in the \nustar data during the dipping intervals. This hard absorption may arise from Compton scattering, modeled with the \texttt{cabs} component, which reduces the flux without altering the spectral shape by scattering photons out of the line of sight.
Overall, we updated the model to \texttt{constant}$\times$\texttt{tbabs}$\times$\texttt{zxipcf}$\times$\texttt{cabs}$\times$\texttt{powerlaw}. 
Since freeing the covering fraction ($f_{\rm cov}$) in the \texttt{zxipcf} component did not significantly improve the fit, we fixed it at 1 during dipping intervals and 0 during non-dipping intervals to minimize parameter coupling.
The equivalent hydrogen column density ($N^{\rm zxipcf}_{\rm H}$) and ionization parameter ($\log\xi$) were allowed to vary during the dipping intervals. 
Additionally, we set the equivalent hydrogen column density in the \texttt{cabs} component, $N^{\rm cabs}_{\rm H}$, to $1.21 \times N^{\rm zxipcf}_{\rm H}$ during the dipping intervals, and to zero during the non-dipping intervals. The factor 1.21 is adopted from \citep{Stelzer1999,DiazTrigo2012} to account for the number of electrons per hydrogen atom in solar-abundance material. Overall, the free parameters include $N^{\rm tbabs}_{\rm H}$, $N^{\rm zxipcf}_{\rm H}$, $\log\xi$, $\Gamma$, and the normalization of the \texttt{powerlaw} component.
We show three examples of the best-fitting spectra at different epochs in the left panels of Fig.~\ref{fig:SPE_plot.png}.

\begin{figure}[!ht]
 \centering
   \includegraphics[width=\linewidth]{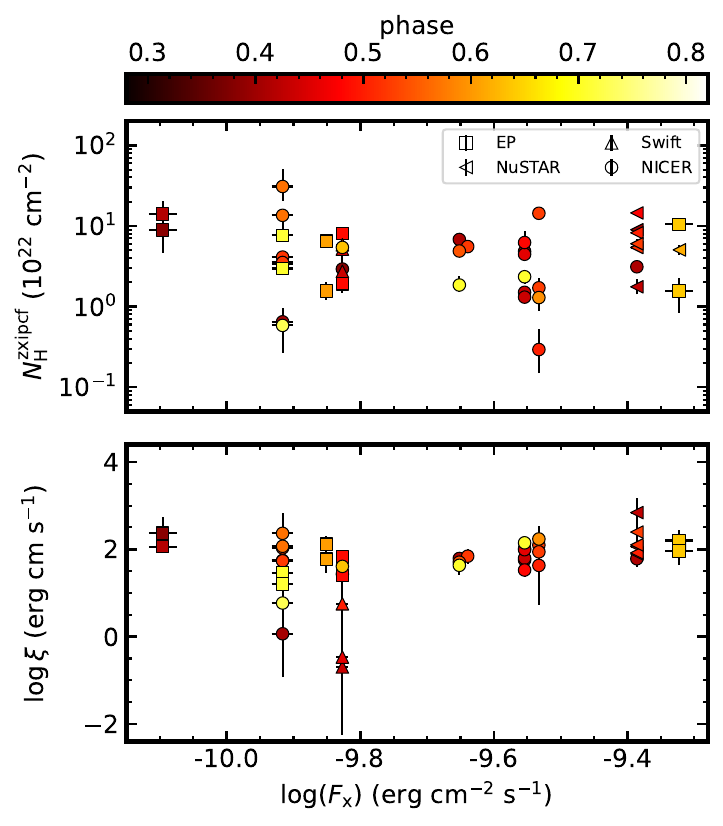}
   \caption{Relationship between the absorption parameters and the unabsorbed flux (0.5--10\,keV). The colorbar denotes the phase range associated with each data point. \textit{Top panel:} $N^{\rm zxipcf}_{\rm H}$ vs. $\log(F_{\rm x})$. \textit{Bottom panel:} $\log\xi$ vs. $\log(F_{\rm x})$.
   }
   \label{fig:rel_L_logxi_nh.png}
\end{figure}

The uncertainties was estimated by Monte Carlo Markov Chain (MCMC) using the \texttt{chain} command in \textsc{xspec}. For each fit, we conservatively used six times the number of free parameters as walkers (e.g. 60 walkers if the model has 10 free parameters) and ran 50,000 steps per walker. The table of best-fitting parameters can be found on GitHub\footnote{\url{https://github.com/zlin-astr/Data\_IGRJ17091\_2025}}. The correlation of $N^{\rm zxipcf}_{\rm H}$ and $\log\xi$ with respect to flux, time, and phase is shown in Figs.~\ref{fig:rel_L_logxi_nh.png} and \ref{fig:SPE_phase_evo_result.png}. The derived absorption parameters exhibit considerable scatter, with $\log\xi\sim1$--3\,\xiunit and $N^{\rm zxipcf}_{\rm H}\sim(1$--$30)\times10^{22}\,\mathrm{cm}^{-2}$. To investigate possible correlations between $N^{\rm zxipcf}_{\rm H}$, $\log\xi$ and the flux, time, and phase, we computed the Spearman rank correlation coefficient ($r_{\rm s}$) using the \texttt{spearmanr} function in the \texttt{SciPy} package \citep{Virtanen2020}. 
To account for parameter uncertainties, we performed 50,000 Monte Carlo simulations by sampling each parameter within its measurement errors. The median $|r_{\rm s}|$ values for all parameter pairs are below 0.3, and their 3$\sigma$ lower bounds are below 0.1. We also examined the correlation between flux and individual phase intervals (e.g., 0.35--0.45, 0.45--0.60, and 0.60--0.75), finding median $|r_{\rm s}|$ values below 0.45 with similarly 3$\sigma$ lower bounds ($<$0.1). The associated median p-values are all greater than 0.05, indicating no statistically significant monotonic correlations.

Given the definition of the ionization parameter \citep{Tarter1969},
\begin{equation}
\xi = \frac{L_{\rm X}}{n_{\mathrm{H}} R^2} = \frac{L_{\rm X}}{N_{\mathrm{H}} R} \cdot \frac{\Delta R}{R}
\end{equation}
where $n_{\rm H}$, $R$, and $\Delta R$ represent the number density, the distance of the absorber from the X-ray source, and its thickness, respectively. The lack of correlation in $N^{\rm zxipcf}_{\rm H}$ and $\log\xi$ as $L_{\rm X}$ decreases by one order of magnitude suggests that either $\Delta R$ increases, $R$ decreases, or both. However, the periodicity of the absorber implies a constant $R$, leaving an increase in $\Delta R$ as the only viable explanation.

Furthermore, we detected a prominent absorption line centered around 7.2\,keV during certain non-dipping intervals,  which is likely associated with highly ionized disk winds rather than the absorber responsible for the dipping episodes. This feature is discussed in detail in Appendix~\ref{sec:Appendix_Xray}.

\subsection{Multi-wavelength Analysis} \label{sec:Multiwavelength_ana}

\subsubsection{optical counterpart}

The optical light curve of \igr is shown in the bottom panel of Fig.~\ref{fig:lc.png}. To explore possible correlations between X-ray and optical variability, we converted the optical magnitudes to fluxes and detrended the long-term decay using an exponential model. We then applied the same LSP analysis to the detrended optical light curves, as shown in the top panels of Fig.~\ref{fig:LSP_CCF_optical.png}. A spike is seen at 2.84\,days with moderate significance ($>2\sigma$). This weaker detection in the optical data may result from limited coverage and/or X-ray reprocessing in the absorber.

Additionally, we computed the discrete cross-correlation function (CCF; \citealt{Sun2018}) between each detrended optical light curve and the detrended \nicer light curve (on a GTI basis). No significant correlation or anti-correlation was found within a lag range of $\pm10$\,day, as shown in the bottom panels of Fig.~\ref{fig:LSP_CCF_optical.png}.

\subsubsection{radio counterpart}
We calculated the in-band photon index ($F \propto \nu^{\alpha}$) using the three 
ATCA observations and obtained $\alpha$ evolving from $-0.16 \pm 0.02$ to $-0.67 \pm 0.04$ within six days (see Tab.~\ref{tab:radio_data}).
These results suggest that the radio source exhibits an optically thin spectrum \citep[][]{Fender2004,Fender2006}, with the peak frequency progressively shifting to lower values as the luminosity decreases.

\section{Discussion} \label{sec:Discussion}
In this work, we investigate the multi-wavelength emission of \igr during its hard-state-only outburst in 2025. For the first time, we identify quasi-periodic X-ray dipping behavior in the system, with a period of 2.83$\pm0.07$\,days and individual dips lasting approximately one day. Our X-ray spectral analysis shows that these dips are primarily caused by absorption and scattering from weakly ionized material along the line of sight. Neither the equivalent hydrogen column density nor the ionization parameter of the absorber shows any significant correlation with time, orbital phase, or luminosity.
The optical light curves show a tentative signal at the same period as the X-ray dips, albeit with a significance just above 2$\sigma$. 
Meanwhile, our radio observations, limited to the early phase of the outburst, reveal a transition from a flat to a steep radio spectrum.
In this section, we discuss the possible physical origins of these dips in the broader context of LMXBs and explore their implications for the physical properties of \igr.

\subsection{Possible origins of the quasi-periodic dips}\label{sec:Discussion_ori}
In Section~\ref{sec:Timing_Analyize}, we showed that the periodic dips likely developed gradually during the decay phase of the 2025 outburst. Notably, \igr also exhibited dips during its 2011 outburst \citep{Pahari2013}, characterized by durations of $\sim$16\,s and recurrence times of $\sim$120\,s. These timescales are significantly shorter than those of the dips analyzed in this work, suggesting a different physical origin.
We examined archival light curves from \nicer, \nustar, \swift, and \textit{RXTE}/PCA, but found no evidence of prominent dips similar to those observed during the 2025 outburst.

Periodic X-ray dips in X-ray binaries (XRBs) have been extensively studied and can arise from a variety of physical mechanisms. These include absorption by asymmetric structures of the accretion disk or associated accretion streams \citep[e.g.,][]{White1982,Frank1987,Trigo2006,Pasham2013,Buisson2021,Raman2018,Jana2022}, shallow eclipses by the donor star \citep[e.g.,][]{Hellier1989,Schmidtke1993,Mukai2003}, precession of tilted or warped disks \citep[e.g.,][]{Wijers1999,Ogilvie2001,Ogilvie2002,Kotze2012,Shaw2013,Martin2023}, and obscuration by circumbinary material \citep[e.g.,][]{Artymowicz1994,Artymowicz1996,Shi2012,DOrazio2013}.

In the scenario where obscuration arises from asymmetric structures in the outer accretion disk \citep{White1982, Frank1987}, LMXBs undergoing Roche-lobe overflow transfer mass from the donor star via a stream that impacts the outer edge of the accretion disk. This interaction creates a dense, vertically extended, and azimuthally asymmetric structure, commonly referred to as the ``bulge", at the stream-disk impact site. As the binary system orbits, this non-axisymmetric bulge periodically crosses the observer’s line of sight, giving rise to quasi-periodic dips in the X-ray light curve that reflect the orbital period of the system.

A detailed study of six LMXBs with periodic dips by \citet{Trigo2006} showed that the ionization parameter of the absorbing material typically lies in the range $\log\xi = 2$–4\,\xiunit. The absorber properties observed in \igr during the dipping episodes are consistent with these findings. In this scenario, variations in the mass accretion rate, mass transfer rate, or the ionization state of the bulge can all influence the formation and geometry of the disk bulge \citep{Armitage1996, Armitage1998}, leading to structural differences between outbursts. Consequently, the sudden appearance of dips in \igr during its 2025 outburst remains an open question.

While the disk bulge scenario provides a plausible explanation for the observations, we briefly consider alternative scenarios and explain why they are less likely in the case of \igr:

Partial obscuration by the donor star can also produce periodic dips in X-ray light curves. Since the dips observed in \igr span approximately 0.36 in phase, this implies an eclipse fraction of $\Delta_{\rm ecl} = 0.36$. We can then estimate the donor star radius ($R_{\rm d}$) relative to its Roche lobe radius ($R_{\rm L,d}$) using the following relations \citep{Chanan1976, Eggleton1983, Porquet2005}:
\begin{align}
\sin i \cos(\pi \Delta_{\mathrm{ecl}}) &= \left[1 - \left({R_{\rm d}}/{a}\right)^2 \right]^{1/2} \\
\frac{R_{\rm L,d}}{a} &= \frac{0.49 q^{2/3}}{0.6 q^{2/3} + \ln(1+q^{1/3})}
\label{equ:RLd}
\end{align}
where $i$ is the inclination angle, $a$ is the binary separation, and $q = M_{\rm d}/M_{\rm c}$ is the mass ratio between the donor star mass ($M_{\rm d}$) and the compact object mass ($M_{\rm c}$). For $\Delta_{\rm ecl} = 0.36$, the derived $R_{\rm d}$ would exceed $1.5R_{\rm L,d}$, resulting in an unphysical configuration. Moreover, such dips would be expected to occur more persistent and should not have been missed in the early \nicer observations in 2025. We therefore do not favor this scenario.

Alternative scenarios, such as the precession of a tilted or warped accretion disk under the tidal forces from the donor star \citep{Ogilvie2001,Ogilvie2002} could in principle produce periodic modulations in optical and X-ray light curves. Such precession could occur over timescales of a few to several hundred days \citep[see, e.g.,][]{Kudryavtsev1989,Homer2001,Kotze2012}, often leading to gradual, sinusoidal variations in X-ray luminosity \citep[e.g.,][]{Kotze2012,Shaw2013}. In the case of \igr, as shown by the X-ray light curve, the dips are sharp, intermittent, and exhibit varying profiles. Outside of the dipping intervals, the light curve remains relatively flat. Thus, the observed 2.84-day period is unlikely associated with disk precession.

Circumbinary disks, which typically reside at large distances from the central binary \citep{Lau2019,Lai2023}, could in principle partially obscure the inner emission and produce absorption features in the light curves.
To explain the appearance of dips in \igr exclusively during the 2025 outburst, one would need to invoke a scenario in which a circumbinary disk gradually moved into the observer’s line of sight, for example, through precession. However, such a scenario would likely lead to a gradual increase in the phase duration of the dips over time. This expected behavior is inconsistent with our observations, as the dipping interval with a phase duration of 0.36 was already established shortly after the dipping feature emerged. Therefore, we do not favor this scenario either.

\subsection{Physical implication of the periodic dips}
In the following, we further explore the implications of the observed dips under the assumption that the dip period corresponds to the binary orbital period.

First, the presence of absorption dips implies a relatively high inclination angle, as the observer’s line of sight must intersect vertically extended structures at the outer edge of the accretion disk. According to \citet{Frank1987}, X-ray binaries that exhibit dipping behavior typically have orbital inclinations between $60^{\circ}$ and $75^{\circ}$. At sufficiently high inclinations, eclipses may occur, whereas at lower inclinations, the observer's line of sight typically does not intersect the outer disk rim, and dips are less likely to be detected. However, reflection modeling estimates the inclination of \igr at $20^{\circ}$--$45^{\circ}$ \citep{Xu2017, Wang2018, Wang2024}; all these studies favor a value considerably lower than the inclination suggested by the dipping behavior.
Similar discrepancies between inclinations derived from reflection modeling and those obtained from dynamical measurements have been reported in other XRBs, such as XTE~J1550--564 \citep{Connors2019}. These differences may be explained by a warped accretion disk, potentially caused by a misalignment between the black hole spin axis and the binary orbital axis \citep[e.g.,][]{Bardeen1975, Natarajan1998, Liska2018, Liska2021}.

Based on the optical brightness of \igr detected during its quiescent state \citep[r$^\prime\sim20.8$ and i$^\prime\sim19.5$;][]{ATel17065}, and adopting a distance of $<25$\,kpc \citep{Altamirano2011}, the donor star's mass is estimated to be less than approximately $10\,M_{\odot}$. Otherwise, the blackbody emission in the r$^\prime$ and i$^\prime$ bands would be brighter than that observed in the quiescent state.
Adopting a typical compact object mass of $M_{\rm c} = 1$--$15\,M_{\odot}$ and a donor mass of $M_{\rm d} = 0.1$--$10\,M_{\odot}$, we estimate the binary separation to be $a = 10$--$30\,R_{\odot}$ for an orbital period of $P_{\rm orb} = 2.84$\,days, based on Kepler’s third law:
\begin{align}
    a = \left[ \frac{G (M_c + M_d) P_{\rm orb}^2}{4\pi^2} \right]^{1/3}.
\end{align}
Using equation equation~(\ref{equ:RLd}), the Roche lobe radius is determined to be $R_{\rm L,d}=2$--$10\,R_{\odot}$. Assuming stable mass transfer occurs when the donor star fills its Roche-lobe (i.e., $R_{\rm d} = R_{\rm L,d}$), the corresponding donor radius across the considered mass range ($M_{\rm d} = 0.1$--$10\,M_\odot$) exceeds the empirical radii expected for main-sequence stars \citep{Eker2018}. This suggests that the donor has likely evolved into a giant star. Moreover, giant stars typically have radii exceeding $10\,R_{\odot}$, which would overfill the estimated Roche lobe in this system. This suggests that the donor in \igr is likely a (partially) stripped giant, an evolved star that has lost part of its envelope, similar to the donors observed in V404 Cyg \citep{King1993} and XTE J1550--564 \citep{Orosz2011a}.

We also independently verified this possibility by evolving a \texttt{MESA} model \citep{Paxton2011, Paxton2013, Paxton2015, Paxton2018, Paxton2019, Jermyn2023} of a $1.2\,M_\odot$ main-sequence donor in a binary with a $10\,M_\odot$ black hole (treat as a point mass) and an initial orbital period of $\sim2.8$ days. The system does not initiate stable mass transfer until the donor radius expands to $\sim 4\,R_\odot$, which occurs well beyond core hydrogen exhaustion ($\sim 1.8\,R_\odot$ for solar metallicity). This supports the conclusion that the donor in \igr has likely evolved into a giant star. Further optical and infrared observations during the quiescent state would help confirm this interpretation. For instance, orbital modulation in the optical/infrared light curves, caused by asymmetric geometry of the Roche-lobe-filling donor, should exhibit the same period as the X-ray dips \citep[e.g.,][]{Orosz2007,Orosz2011}. Additionally, infrared spectroscopy could help constrain the spectral type of the donor star \citep[e.g.,][]{Casares1994,Greiner2001}.

\newpage

\begin{acknowledgments}

We thank the anonymous referee for the constructive comments, which have helped improve the manuscript. We thank Ileyk El Mellah for the useful discussion. This research was supported by the National Natural Science Foundation of China (NSFC) under grant No. 12588202; by the New Cornerstone Science Foundation through the New Cornerstone Investigator Program and the XPLORER PRIZE; by the Strategic Priority Program of the Chinese Academy of Sciences under grant No. XDB0550203, and by the China-Chile Joint Research Fund (CCJRF No.2301) and the Chinese Academy of Sciences South America Center for Astronomy (CASSACA) Key Research Project E52H540301. ZG is funded by ANID, Millennium Science Initiative, AIM23-001.

\end{acknowledgments}

\bibliography{main}{}
\bibliographystyle{aasjournal}

\appendix
\renewcommand{\thefigure}{\hbAppendixPrefix\arabic{figure}}
\numberwithin{figure}{section}

\section{Detailed information on optical variability}  \label{sec:Appendix_Opt}

\begin{figure*}[!ht]
 \centering
   \includegraphics[width=\linewidth]{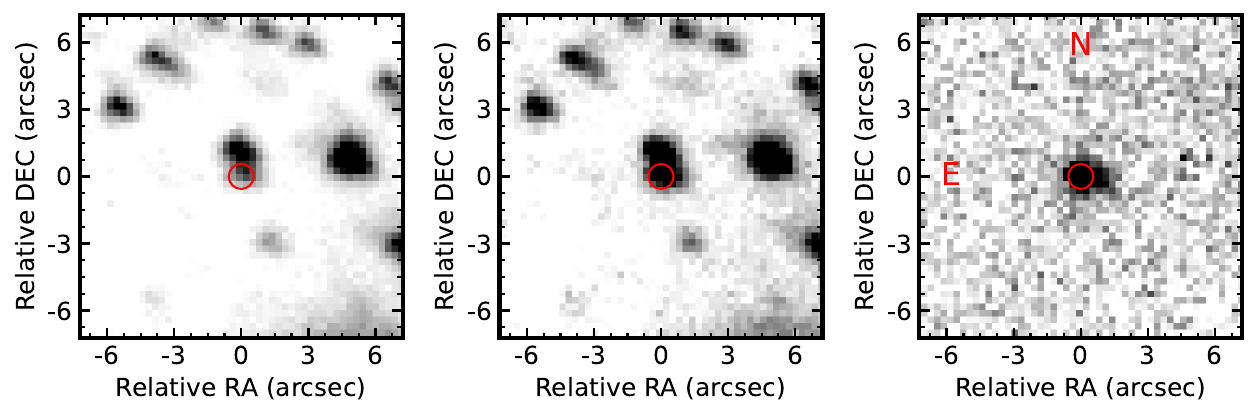}
   \caption{\textit{Left panel}: SOAR image taken at 125\,days. \textit{Middle panel}: SOAR image taken at 85\,days. \textit{Right panel}: Difference image between 85 and 125\,days. Images were observed in the $i^\prime$-band. The red circle, marking the position of \igr, is centered at R.A. = $257.28166^\circ \pm 0.26\arcsec$ and Decl. = $-36.40711^\circ \pm 0.14\arcsec$ (J2000), as determined from the ATCA measurements.
   }
   \label{fig:SOAR_diff_image.png}
\end{figure*}

Based on the SOAR imaging data, \igr can be partially resolved from a nearby contaminating source. To robustly determine whether the observed optical variability originates from our target, we performed image subtraction using the \texttt{HOTPANTS} package \citep{Becker2015}, which implements the algorithm developed by \citet{Alard1998}. The resulting difference image (see Fig.~\ref{fig:SOAR_diff_image.png}) clearly shows that the variability arises from \igr, while the nearby source remains relatively stable.

\begin{figure*}[!ht]
 \centering
   \includegraphics[width=\linewidth]{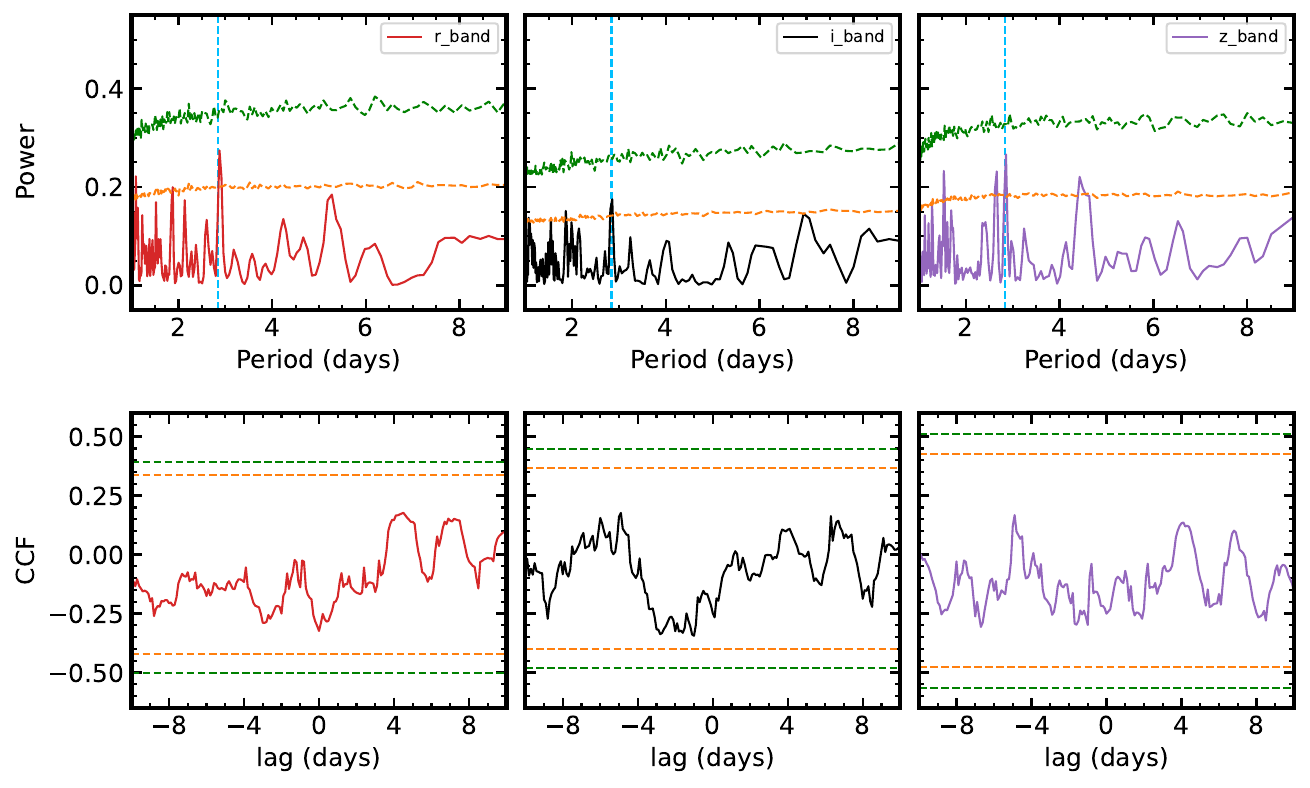}
   \caption{\textit{Top panels}: LSP of the detrended optical light curve in the $r^\prime$ (left), $i^\prime$ (middle), and $z^\prime$ (right) filters.
   \textit{Bottom panels}: CCF between the detrended \nicer and optical light curves. The orange and green dashed lines indicate the $95\%$ ($2\sigma$) and $99.73\%$ ($3\sigma$) significance level, respectively. The vertical blue dashed line marks the period of 2.84\,days. 
   }
   \label{fig:LSP_CCF_optical.png}
\end{figure*}

In this section, we present the LSPs of the detrended optical light curves, together with the CCFs between each detrended optical light curve and the detrended \nicer light curve, in Fig.~\ref{fig:LSP_CCF_optical.png}.

\section{Parameters of the absorption dips} \label{sec:Appendix_dips}

\begin{figure}[!ht]
 \centering
   \includegraphics[width=\linewidth]{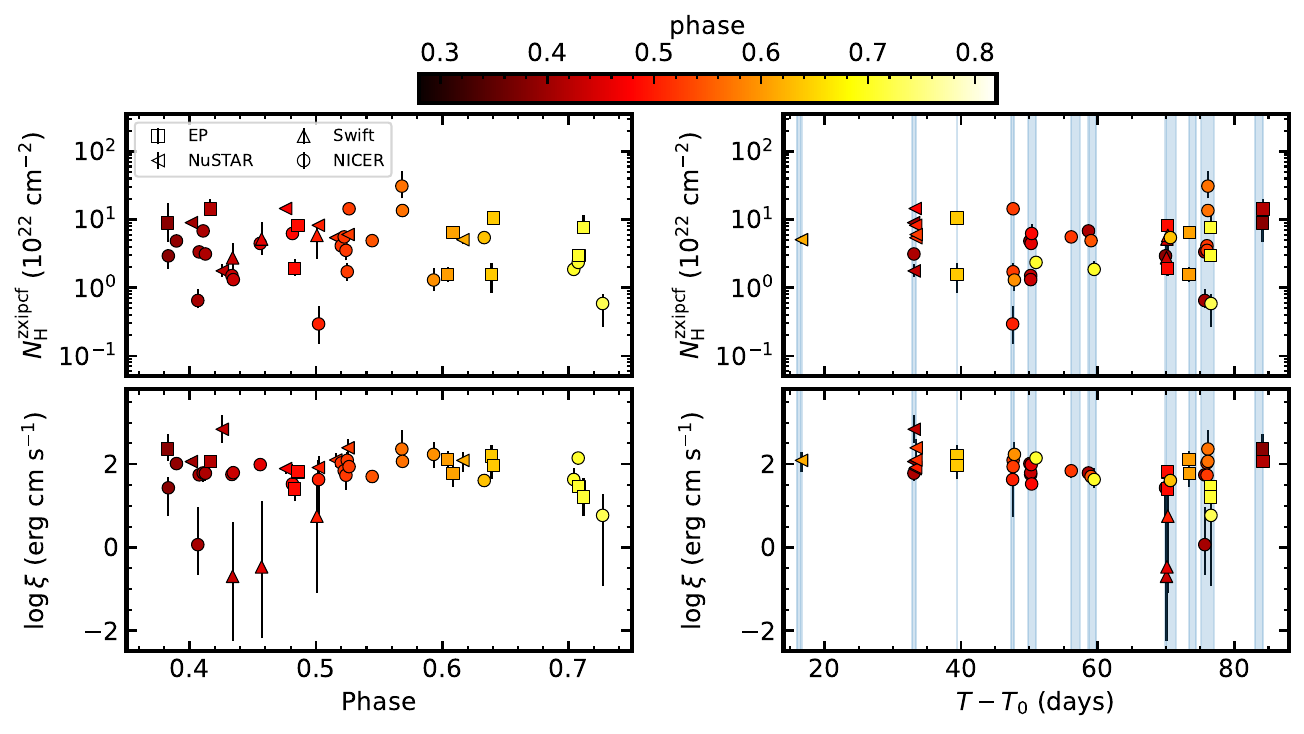}
   \caption{\textit{left panels:} The best-fitting absorption parameters shown in phase space. The colorbar represents the phase range corresponding to each data point.
   \textit{Right panels:} The best-fitting absorption parameters shown in time space. The blue shaded region indicates the time coverage for the joint fit episodes. 
   }
   \label{fig:SPE_phase_evo_result.png}
\end{figure}

In this section, we present the evolution of the absorption parameters in phase and time space, as shown in Fig.~\ref{fig:SPE_phase_evo_result.png}.

\section{Potential disk wind during the hard state} \label{sec:Appendix_Xray}

\begin{figure*}[!ht]
 \centering
   \includegraphics[width=0.5\linewidth]{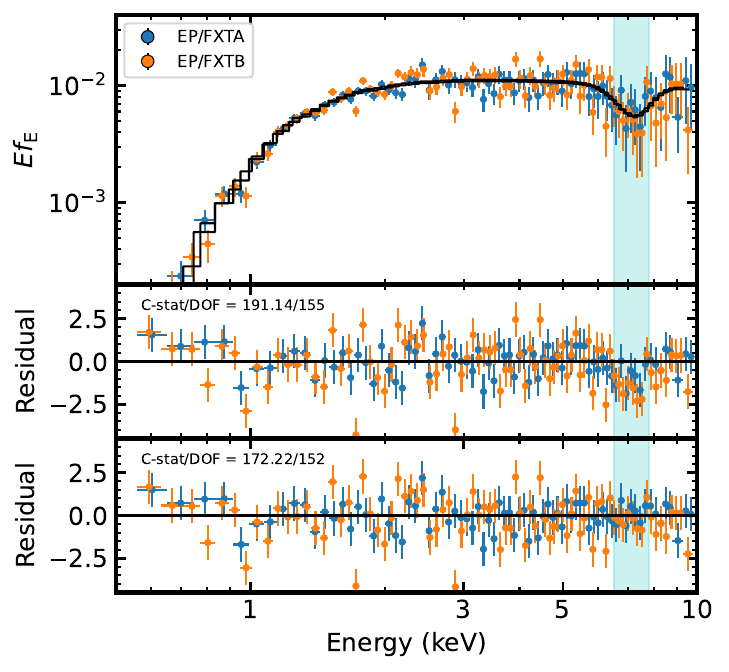}
   \caption{The \ep spectrum observed at 81.3\,days (non-dipping episode). The cyan shaded region highlights a prominent absorption feature centered at 7.2\,keV. The bottom two panels show the residuals of the fit without and with an additional \texttt{gabs} component in the continuum model (\texttt{constant}$\times$\texttt{tbabs}$\times$\texttt{powerlaw}). The fit statistics (C-stat/DOF) for both models are showed in the upper left corner.
   }
   \label{fig:EP_553.png}
\end{figure*}

In several individual X-ray spectra of \igr, an absorption feature around 7\,keV occasionally appears. The observation by \ep at 81.3\,days shows the most prominent example (see Fig.~\ref{fig:EP_553.png}). We convolve a \texttt{gabs} component with the continuum, which significantly improves the fit statistics by $\Delta$C-stat=19 for three additional DOFs. The best-fitting centroid energy is $7.2_{-0.2}^{+0.3}$\,keV, consistent with the energy range of Fe K edge or blue-shifted Fe K$\alpha$ lines \citep[e.g.,][]{Furst2011,Koljonen2020}. 
Since the observation was made during the non-dipping episodes, this feature likely has a different origin from the other absorption components observed and possibly associated with disk winds, as seen in previous outbursts in both the hard \citep[e.g.,][]{Wang2018} and soft states \citep[e.g.,][]{Wang2024}. Hydrodynamic simulations suggest that, during the hard state of XRBs, disk winds may be launched with higher velocities and degrees of ionization compared to those observed in the soft state \citep{Higginbottom2020}. A detailed analysis of the absorption lines will be provided in a separate work.

\end{document}